\documentclass[fleqn,usenatbib]{mnras}

\usepackage{newtxtext,newtxmath}
\usepackage[T1]{fontenc}
\usepackage{ae,aecompl}

\usepackage{graphicx}	% Including figure files
\usepackage{amsmath}	% Advanced maths commands
\usepackage{booktabs}
\usepackage{caption}
\usepackage{enumitem}

\usepackage[dvipsnames]{xcolor}
\usepackage[decimalsymbol=., expproduct=times, separate-uncertainty=true, multi-part-units=single]{siunitx} % SI units 
\DeclareSIUnit\erg{erg}

\newcommand{\MSun}{$\mathrm{M}_{\odot}\mathrm{\ }$} 
\newcommand{\MSunns}{$\mathrm{M}_{\odot}$} 
\newcommand{\MJup}{$\mathrm{M}_{\mathrm{Jup}}\mathrm{\ }$} 
\newcommand{\MEarth}{$\mathrm{M}_{\oplus}$} 

\title[Effects of stellar density on photoevaporation of circumstellar discs]{Effects of stellar density on the photoevaporation of circumstellar discs}

\author[Concha-Ram\'irez et al.]{
Francisca Concha-Ram\'irez,$^{1}$\thanks{fconcha@strw.leidenuniv.nl}
Martijn J. C. Wilhelm,$^{1}$
Simon Portegies Zwart,$^{1}$\newauthor
Sierk E. van Terwisga,$^{2}$
Alvaro Hacar$^{1, 3}$\\
$^{1}$
Leiden Observatory, Leiden University, PO Box 9513, 2300 RA Leiden, The Netherlands\\
$^{2}$
Max-Planck-Institut für Astronomie, Königstuhl 17, D-69117 Heidelberg, Germany\\
$^{3}$
University of Vienna, Department of Astrophysics, Türkenschanzstrasse 17, 1180 Vienna, Austria
}

\date{Accepted XXX. Received YYY; in original form ZZZ}

\pubyear{2020}

\begin{document}
\label{firstpage}
\pagerange{\pageref{firstpage}--\pageref{lastpage}}
\maketitle

\begin{abstract}
Circumstellar discs are the precursors of planetary systems and develop shortly after their host star has formed. In their early stages these discs are immersed in an environment rich in gas and neighbouring stars, which can be hostile for their survival. There are several environmental processes that affect the evolution of circumstellar discs, and external photoevaporation is arguably one of the most important ones. Theoretical and observational evidence point to circumstellar discs losing mass quickly when in the vicinity of massive, bright stars. In this work we simulate circumstellar discs in clustered environments in a range of stellar densities, where the photoevaporation mass-loss process is resolved simultaneously with the stellar dynamics, stellar evolution, and the viscous evolution of the discs. Our results indicate that external photoevaporation is efficient in depleting disc masses and that the degree of its effect is related to stellar density. We find that a local stellar density lower than 100 stars pc$^{-2}$ is necessary for discs massive enough to form planets to survive for \SI{2.0}{Myr}. There is an order of magnitude difference in the disc masses in regions of projected density 100 stars pc$^{-2}$ versus $10^4$ stars pc$^{-2}$. We compare our results to observations of the Lupus clouds, the Orion Nebula Cluster, the Orion Molecular Cloud-2, Taurus, and NGC 2024, and find that the trends observed between region density and disc masses are similar to those in our simulations.
\end{abstract}


%%%%%%%%%%%%%%%%%%%%%%%%%%%%%%%%%%%%%%%%%%%%%%%%%%

%%%%%%%%%%%%%%%%% BODY OF PAPER %%%%%%%%%%%%%%%%%%

\section{Introduction}
Circumstellar discs are the reservoirs of gas and dust that surround young stars and have the potential to become planetary systems. Their evolution will determine the time and material available to form planets. Studying the evolution of circumstellar discs can then help us understand planet formation and the diversity of observed planetary systems.

These circumstellar discs develop almost immediately after star formation, as a direct consequence of the collapse of a molecular cloud and angular momentum conservation \citep{williams2011}. Their surroundings are rich in gas and neighbouring stars, which can be hostile to the discs and affect their evolution in different ways. In environments with high stellar densities, dynamical encounters with nearby stars can truncate the discs \citep[e.g.][]{pfalzner2006, vincke2015, portegieszwart2016, bhandare2019}. Face-on accretion of gas onto the circumstellar discs can cause them to shrink and increase their surface densities \citep[e.g.][]{wijnen2016, wijnen2017}. Feedback from processes related to stellar evolution, such as stellar winds and supernovae explosions, can also truncate, tilt, or completely destroy the discs \citep{pelupessy2012, close2017, portegieszwart2018}. The presence of bright, massive stars in the vicinity of circumstellar discs can heat their surface enough to evaporate mass from them. This process, known as external photoevaporation, is arguably one of the most important environmental mechanisms in depleting mass from young circumstellar discs, and its effects seem to greatly outperform that of other means for disc truncation \citep[e.g.][]{hollenbach2000, adams2004, guarcello2016, facchini2016, winter2018a, haworth2019, winter2020a, haworth2020a}. 

The effects of external photoevaporation have been identified in observational surveys of young stellar objects in star-forming regions. Proplyds --cometary tail-like structures formed by ionized, evaporating discs-- have been observed in particular in dense regions of the Orion nebula \citep{odell1994, odell1998, vicente2005, eisner2006, mann2014, kim2016}. Surveys of protoplanetary discs in star-forming regions show that discs closer to bright stars are less massive than their counterparts in sparser regions \citep{fang2012, mann2009, mann2014, ansdell2017, vanterwisga2020}, suggesting that discs in the vicinity of these stars are strongly affected by their environment. Disc fractions (the number of young stellar objects around which dust is detected, over the total number of objects) and disc mass distributions in younger and less dense star-forming regions, such as Lupus and Taurus, are statistically indistinguishable from each other in terms of disc mass distributions. The average disc mass in these regions is higher than in the Orion Nebula Cluster (ONC) \citep{ansdell2016, eisner2008, eisner2018, vanterwisga2019}, which is a much denser environment. 

The extent of the effects of radiation in depleting disc mass depends on the proximity to bright stars. As such, the effects of external photoevaporation are important in clustered environments, and different theoretical models have been developed to study the process in that context. \citet{scally2001} model the dynamics of a cluster with 4000 stars, with discs of radii \SI{100}{au} which remain constant throughout the simulation. The mass loss due to photoevaporation is calculated in post processing, by keeping track of the radiation to which each star is exposed during the simulation. Their results show that external photoevaporation is important in depleting disc mass in regions similar to the centre of the ONC. A different approach is taken by \citet{adams2006} and \citet{fatuzzo2008}, who model the dynamics of star clusters of different sizes and derive the background FUV radiation for each simulation. They then estimate the photoevaporation mass loss rates of the discs depending on the average background radiation that hey have been exposed to. \citet{adams2006} model star clusters with 100, 300, and 1000 stars and find that external photoevaporation is only important for disc radii larger than \SI{30}{au}, due to the low average background UV exposure. Furthermore, models of single, externally illuminated discs show that the supersonic flows caused by far-ultraviolet (FUV) photons heating the disc surface can explain the observed proplyd shapes \citep[e.g.][]{richling1997, johnstone1998, storzer1999, adams2004}.

In more recent work, \citet{winter2018a} and \citet{winter2020} use a statistical approach to model background FUV radiation fields in regions of different stellar number density. \citet{winter2020} find that 90\% of circumstellar discs are destroyed by external photoevaporation within \SI{1.0}{Myr} in a region comparable to the Central Molecular Zone of the Milky Way, and that the effects of photoevaporation are particularly destructive for discs around low mass stars ($\mathrm{M}_* < 0.3 \mathrm{M}_{\odot}$). For regions similar to the solar neighbourhood (surface density $\Sigma_0 = 12 \mathrm{\ M}_{\odot} \mathrm{\ pc}^{-2}$) they find a mean dispersal timescale of $\sim\SI{3.0}{Myr}$. Similar results are obtained by \citet{nicholson2019}, who calculate mass loss rates in N-body simulations using the same prescriptions as \citet{scally2001} and find that, in regions with high degree of substructure (density $\sim 100$ \MSun pc$^{-3}$), 50\% of the discs with initial radii $\geq \SI{100}{au}$ are destroyed by external photoevaporation within \SI{1}{Myr}. In regions of lower densities ($\sim 10$ \MSun pc$^{-3}$), half of the discs are destroyed within \SI{2}{Myr}.

% Move this to discussion
%Photoevaporation quickly destroying circumstellar discs is not the only thing that shows that planets must start forming early. The 'missing-mass problem', an observational discrepancy in which protoplanetary disc masses are lower than the masses of known rocky planetary systems, also suggests that planet formation is already ongoing in discs of ages as small as $1.0 - \SI{3.0}{Myr}$ \citep{greaves2010, williams2012, najita2014, manara2018}, or even earlier than $\SI{0.5}{Myr}$ \citep{tychoniec2020}. Both the short lifetimes of circumstellar discs even in low radiation fields \citep{facchini2016} and the missing-mass problem point to planet formation starting shortly after circumstellar discs form. 

In \citet{concha-ramirez2019a} (hereafter Paper I) we presented a new numerical implementation which allows for self-contained simulations of external photoevaporation in clustered environments. Photoevaporation process is solved simultaneously with the stellar dynamics (including disc encounters and truncations), stellar evolution, and viscous spreading of the circumstellar discs. This causes disc masses, sizes, and column densities to vary in time, and the mass loss rate of the discs is calculated accordingly. The results of the simulations in Paper I show that external photoevaporation is efficient in destroying circumstellar discs on a relatively short timescale. For regions of stellar densities $\sim 100 \mathrm{M}_{\odot} \mathrm{pc}^{-3}$, around $80\%$ of discs have evaporated within \SI{2.0}{Myr}. Between 25\% and 60\% of the discs, depending on region density, are destroyed within the first \SI{0.1}{Myr}. We argue that the rapid decrease in disc mass is dominated by external photoevaporation, rather than dynamical truncations, and that the former mechanism constrains the time available for planet formation.

Observational and theoretical evidence suggest that the local stellar density is a key factor in the survival of circumstellar discs and in their eventual observed mass distributions. Understanding disc mass and size distributions in young star clusters is therefore paramount for understanding planet formation and evolution. Here we use the model developed in Paper I to determine for which range of stellar densities the effects of photoevaporation are most efficient. We perform simulations of circumstellar discs embedded in star clusters and explore a parameter space of stellar densities spanning five orders of magnitude. The clusters are evolved for \SI{2.0}{Myr} and we investigate the final mass distributions of the disc population. We compare our simulation results to observed dust masses of young stellar objects in the Lupus clouds \citep{ansdell2016, ansdell2018}, the Orion Nebula Cluster \citep{mann2014, eisner2018}, the Orion Molecular Cloud-2 \citep{vanterwisga2019}, the Taurus region \citep{andrews2013}, and NGC 2024 \citep{getman2014, vanterwisga2020}.

\section{Model}
We use the Astrophysical Multipurpose Software Environment, AMUSE\footnote{\url{http://amusecode.github.io}} \citep{portegieszwart2009, portegieszwart2013}, to bring together codes for viscous disc evolution, stellar dynamics, and stellar evolution, along with an implementation of external photoevaporation. The setup and models used for the simulations in this paper are the same as in Paper I. In the present work we perform simulations spanning a larger range of stellar densities. 

Here we present a summary of the implementation used for the simulations. For a detailed explanation of the model the reader should refer to Paper I. All the code developed for the simulations, data analyses, and figures of this paper is available online\footnote{\url{https://doi.org/10.5281/zenodo.3897171}}.

\subsection{Stars and circumstellar discs}\label{model:discs}
We separate the stars in the simulations into two populations: stars with masses $\mathrm{M}_* \leq 1.9$ \MSunns, and stars with masses $\mathrm{M}_* > 1.9$ \MSunns. The reason for this mass limit is related to the photoevaporation mass loss calculation and further explained in section \ref{model:photoevaporation}. This mass separation is for photoevaporation purposes only and does not influence the dynamical evolution of the stars. All stars with masses $\mathrm{M}_* \leq 1.9$ \MSun are surrounded by a circumstellar disc, while stars with higher masses have no discs and are considered only as generating ionizing radiation. Massive stars are subject to stellar evolution, implemented using the code SeBa \citep{portegieszwart1996, toonen2012} through its AMUSE interface. Stars with discs do not undergo stellar evolution in the simulations. The dynamical evolution of the clusters is implemented using the 4th-order N-body code ph4, incorporated in AMUSE.  

Circumstellar discs are implemented using the Viscous Accretion disc Evolution Resource (VADER) developed by \citet{krumholz2015}. VADER models mass and angular momentum transport on a thin, axisymmetric disc. This allows us to take into consideration the viscous spreading of the discs. Each VADER disc in our simulations is composed of a grid of 100 logarithmically spaced cells between 0.05 and $\SI{2000}{au}$. The discs have a turbulence parameter of $\alpha=\num{5e-3}$.

The initial disc column density follows the standard disc profile by \citet{lynden-bell1974}, with characteristic radius $r_c \approx r_d$ \citep{anderson2013}. To properly calculate the photoevaporation mass loss rate we need to keep track of the outer disc edge (see section \ref{model:photoevaporation}) which we define as the cell closest to a low column density value, $\Sigma_{\mathrm{edge}}$ \citep{clarke2007, haworth2018a}. We set the column density outside $r_d$ to a negligible value $\Sigma_{\mathrm{edge}} = \SI{e-12}{\gram\per\square\cm}$. The initial surface density then takes the form:

\begin{equation}
\Sigma(r, t=0) = 
\begin{cases}
\frac{m_d}{2 \pi r_d \left(1 - e^{-1}\right)} \frac{\exp(-r/r_d)}{r} & \mathrm{\ for\ } r \leq r_d,\\
\\
\SI{e-12}{\gram\per\square\cm} & \mathrm{\ for\ } r > r_d,
\end{cases}
\end{equation}

\noindent
where $r_d$ and $m_d$ are the initial radius and mass of the disc, respectively.

\subsection{External photoevaporation}\label{model:photoevaporation}
External photoevaporation is dominated by far-ultraviolet (FUV) photons \citep{armitage2000, adams2004, gorti2009a}. To model the FUV radiation from the massive stars we pre-compute a relation between stellar mass and FUV luminosity using the UVBLUE spectral library \citep{rodriguez-merino2005}. The obtained FUV luminosity fit is shown in Figure 2 of Paper I. During the simulations we use this fit to obtain the FUV luminosity of each massive star at every time step.

Mass loss due to external photoevaporation is calculated for each disc using the Far-ultraviolet Radiation Induced Evaporation of Discs (FRIED) grid \citep{haworth2018}. The FRIED grid provides a set of pre-calculated, external photoevaporation mass loss rates for discs immersed in radiation fields of varying intensity, from $10 \mathrm{\ G}_0$ to $10^{4} \mathrm{\ G}_0$. The grid spans discs of mass $\sim10^{-4} \mathrm{\ M}_\mathrm{Jup}$ to $10^{2} \mathrm{\ M}_\mathrm{Jup}$, radius from $\SI{1}{au}$ to $\SI{400}{au}$, and host star mass from $0.05 \mathrm{\ M}_{\odot}$ to $1.9 \mathrm{\ M}_{\odot}$. To stay within the limits of the grid, we give all stars with masses $\mathrm{M}_* \leq 1.9$ \MSun a circumstellar disc, and all stars with masses $\mathrm{M}_* > 1.9$ \MSun are considered as only generating radiation.

We calculate the mass loss of every disc as follows. For each disc we begin by calculating its distance to every star of mass $\mathrm{M}_* > 1.9$ \MSun and determining the total radiation that the disc receives from those stars. We do not consider extinction in this calculation. We then use this total radiation and the disc parameters to interpolate a mass loss rate $\dot{\mathrm{M}}$ from the FRIED grid. This $\dot{\mathrm{M}}$ is then used to calculate the total mass lost by the disc in the current time step. Assuming a constant mass loss rate over the time step, the mass is removed from the outer regions of the disc: we advance over the disc cells starting from the outside removing mass from each, until the corresponding amount of mass has been removed. Through this process, mass loss due to photoevaporation results in a decrease of disc mass and disc radius.

In some cases a massive star gets close enough to a disc to enter a photoevaporation regime dominated by extreme ultraviolet (EUV) radiation. In this regime the mass loss is calculated as \citet{johnstone1998}: %We define a distance limit for this case following \citet{johnstone1998}:

%\begin{equation}\label{eq:dmin}
%    d_{min} \simeq 5 \times 10^{17} \left(\frac{\epsilon^2}{f_r \Phi_{49}}\right)^{-1/2} \textrm{r}_{d_{14}}^{1/2} \mathrm{\ cm}
%\end{equation}

%\noindent
%where $f_r$ is the fraction of EUV photons absorbed in the ionizing flow, $\Phi_{49} = \frac{\Phi_i}{10^{49}} \mathrm{s}^{-1}$ is the EUV photon luminosity of the source, and $\epsilon$ is a dimensionless normalizing parameter. The factor $\left(\frac{\epsilon^2}{f_r \Phi_{49}}\right)^{1/2} \approx 4$, and $r_{d_{14}} = \frac{r_d}{10^{14} cm}$ with $r_d$ the disc radius.%, and $5 \times 10^{17} \mathrm{\ cm} \sim 3 \times 10^4 \mathrm{\ au} \sim 0.16 \mathrm{\ pc}$. 
%When the distance $d$ between a disc and a massive star is $d < d_{min}$, EUV photons dominate the radiation and the mass loss is calculated as:

\begin{equation}\label{eq:euv}
    \dot{M}_{EUV} = 2.0 \times 10^{-9} \frac{(1 + x)^2}{x} \epsilon r_{d_{14}} M_{\odot} \mathrm{\ yr}^{-1}
\end{equation}

\noindent
with $x \approx 1.5$ and $\epsilon \approx 3$.

We consider a disc dispersed when its mass drops below $0.03$ \MEarth, based on the non-detection mass limits from \citep{ansdell2016}, or when its mean column density is lower than 1 g cm$^{-2}$ \citep{pascucci2016} (see also Section \ref{caveats}).

\subsection{Initial conditions}\label{ics}
\subsubsection{Star clusters}
We simulate clusters with $10^3$ stars and initial virial radii of 0.1, 0.3, 0.5, 1.0, 2.5, and $\SI{5.0}{pc}$. Stars are initially distributed in a Plummer sphere \citep{plummer1911}. Stellar masses are drawn from a random Kroupa mass distribution \citep{kroupa2001} with upper limit 100 \MSunns. All models start in virial equilibrium (virial ratio $Q=0.5$). No primordial mass segregation, binaries, or higher multiplicity systems are considered.\\

In Table \ref{table:models} we present the models used for this work. The mean number of stars with discs in each simulation is $974.7 \pm 1.7$. The mean mass of the stars with discs is $0.23\substack{+1.66 \\ -0.22} \mathrm{M}_{\odot}$. The mean number of stars generating UV radiation is $25.3 \pm 1.7$. The third column of Table \ref{table:models} shows the mass ranges spanned by these stars.

We evolve each cluster for \SI{2.0}{Myr}. We run each model 6 times, with a different random seed for the mass function and the initial stellar positions and velocities.

\begin{table}
\centering
\def\arraystretch{1.5}
\begin{tabular}{ccccc} 
\toprule
Model name & $\mathrm{R}_\text{vir}$ [pc] & $\overline{\mathrm{M}}_{\mathrm{M}_* > 1.9 \mathrm{M}_{\odot}}$ [\MSunns] & $\overline{\mathrm{N}}_{*\mathrm{B}}$ & $\overline{\mathrm{N}}_{*\mathrm{O}}$\\
\midrule
R0.1 & 0.1 & $6.61\substack{+57.18 \\ -7.57}$ & $23.5 \pm 1.1$ & $2.2 \pm 1.1$\\
R0.3 & 0.3 & $6.62\substack{+81.98 \\ -7.07}$ & $22.5 \pm 2.8$ & $2.5 \pm 0.5$\\
R0.5 & 0.5 & $5.22\substack{+53.54 \\ -4.41}$ & $25.1 \pm 2.5$ & $1.8 \pm 1.1$\\
R1.0 & 1.0 & $5.61\substack{+41.72 \\ -5.56}$ & $22.0 \pm 3.0$ & $1.8 \pm 0.1$\\
R2.5 & 2.5 & $5.94\substack{+46.09 \\ -5.06}$ & $23.8 \pm 7.8$ & $1.8 \pm 1.2$\\
R5.0 & 5.0 & $6.37\substack{+76.43 \\ -5.24}$ & $25.2 \pm 4.3$ & $2.5 \pm 1.5$\\
\bottomrule
\end{tabular}\caption{Simulation models. First column: model name. Second column: initial virial radius, in parsec. Third column: mean mass of radiating stars ($\mathrm{M}_* > 1.9 \mathrm{M}_{\odot}$), in \MSunns. Fourth column: mean number of B type stars. Fifth column: mean number of O type stars. All means are calculated over 6 runs for each model.}\label{table:models}
\end{table}

\subsubsection{Circumstellar discs}
Observations of resolved circumstellar discs suggest they are generally compact, with radii around 20 to $\SI{50}{au}$ \citep{trapman2020, tobin2020}. The initial radii of the discs in our simulations are given by:

\begin{equation}\label{eq:discradius}
r_d(t=0) = R'\left(\frac{M_*}{M_\odot}\right)^{0.5},
\end{equation}

\noindent
where $R'$ is a constant. We choose $R' = \SI{30}{au}$, which yields an initial disc radii distribution between $\sim\SI{5}{au}$ and $\sim\SI{40}{au}$. Initial discs masses are defined as 10\% of the mass of their host star.

%\begin{equation}\label{eq:discmass}
%\mathrm{M}_d(t=0) = 0.1 \mathrm{M}_*    
%\end{equation}

%\noindent
%where $\mathrm{M}_*$ is the mass of the host star. 

\subsection{Model caveats}\label{caveats}
Our model is designed as a controlled experiment to investigate the physical processes going on inside star-forming regions, in particular with regards to external photoevaporation. There are quite a number of assumptions in our simulations which we justify based on previous theoretical work and observations. Below we discuss some of these processes and parameters and the implications they might have.

Star-forming regions are not only rich in stars but also in gas, which can linger for several million years \citep{portegieszwart2010}. Intracluster gas could influence our results in two main ways: first, the presence of gas and its subsequent expulsion in time affect the virial equilibrium and thus the dynamics of the star clusters. Second, gas can absorb some of the FUV radiation coming from bright stars, effectively protecting the discs from external photoevaporation and allowing them to live for longer \citep{winter2019a, ali2019, vanterwisga2020}, therefore giving more time for the planet formation process to occur. The presence of gas can also explain the 'proplyd lifetime problem' observed in the ONC, in which discs not massive enough to survive in the environment of $\theta ^1 C$ Ori are still observed in the region \citep{winter2019a}.

%Another parameter which can alter the amount of radiation received by each disc is its inclination. Depending on its orientation, a disc can be irradiated by a nearby star face on, edge on, or a position in between. This can generate a discrepancy of mass loss in different regions of the disc. We do not give the discs in our model a particular orientation, but consider the effect of inclination averaged out as we take into account radiation coming from massive stars in different directions. 

The FRIED grid was constructed using a 1-dimensional disc model, but later simulations by \citet{haworth2019} show that mass loss rates can increase up to a factor of 3.7 when considering 2-dimensional discs. It is likely then that the mass losses used in this work are only a lower limit for the effects of external photoevaporation.

Internal photoevaporation, the process in which X-Ray and UV photons coming from the host star itself lead to mass loss, is not considered in these simulations. Internal photoevaporation can drive mass loss in the inner regions of the discs \citep[$\sim1 - \SI{10}{au}$ and $\SI{30}{au}$,][]{gorti2009, gorti2009a} and even in outer regions under certain conditions \citep{owen2010, font2004}. However, external photoevaporation is arguably the dominant process in regions $>\SI{10}{au}$ \citep{hollenbach2000, fatuzzo2008}. Our approximation of external photoevaporation removing mass from the outer regions of the disc only is also idealized, since while internal photoevaporation seems to clear the disc at specific radii, FUV photons coming from external sources can heat and evaporate mass from the whole disc surface \citep{adams2004}. 

Our conditions for disc dispersal might be overestimating the number of destroyed discs, particularly for relatively high mass stars. In some cases, a disc of density 1 g cm$^{-2}$ or lower can still be detectable with modern instruments. As the mean mass of stars with discs in our simulations is 0.23 \MSun (corresponding to an initial disc mass of 24 \MJup), and as most of the massive discs in our simulations survive until the end (see section \ref{results:discfractions}), we consider this possible underestimation of surviving discs to be within the uncertainty of our simulations and to not affect our results.

Our prescription of the EUV mass loss rate (Eq. \ref{eq:euv}) is taken from \citet{johnstone1998} and corresponds to a thick photodissociation region (PDR). In practice, in EUV dominated fluxes the PDR is expected to be thin. This can change the value of the EUV mass loss rate $\dot{M}_{EUV}$ calculated in each time step. However, since our focus is on FUV photoevaporation, and since discs are only in the EUV regimes for a short time in our simulation, this hardly affects our mass loss rate calculations.

Regarding disc masses, it is generally accepted that a 100:1 gas-to-dust mass ratio defines the composition of circumstellar discs. However, several authors have pointed out that this value might change across discs and in time \citep{williams2014, manara2020}. This can lead to observed disc dust masses being greatly underestimated \citep{manara2018}. New models of externally irradiated, evaporating discs by \citet{haworth2018a} show that considering grain growth can lead to less dust being lost through external photoevaporation, and thus to the dust:gas ratio increasing in time. A more careful implementation of the separate dust and gas components in a disc can help to overcome this problem.

The distribution of stars in a Plummer sphere is an idealized geometry. Star-forming regions have complex configurations and can present fractal structures, filaments, and other regions of increased surface density \citep[e.g.][]{scalo1990, elmegreen1996, elmegreen2000, bate2010, hacar2013, chevance2020, krause2020}. The simulations carried out for this work represent only local densities, but for improved analyses of disc survival in star-forming regions it is important to consider different spatial distributions.

\section{Results}

\begin{figure*}
  \includegraphics[width=0.85\linewidth]{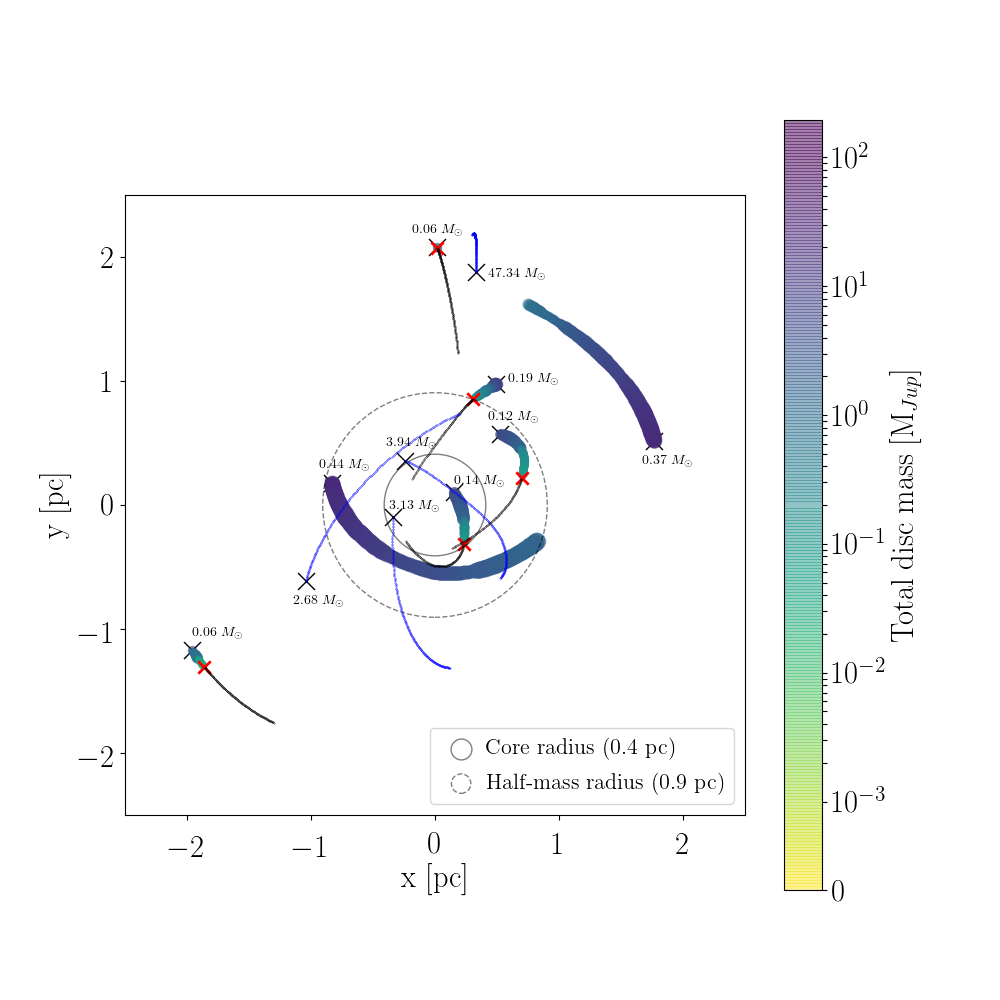}
  \caption{Example of cluster evolution for a realisation of the R1.0 model. Black crosses mark the position of the stars at the beginning of the simulation, and the label next to them shows the stellar mass. The sizes of the large, coloured points are proportional to the disc radii, and their colour indicates the total disc mass. The red crosses, when present, show the moment when a disc is dispersed. The thin black lines that follow a red cross indicate the continuing orbit of the star, which keeps moving through the region after its disc has been evaporated. The trajectories of some massive, radiating stars are shown in thin blue lines.}
  \label{fig:tracks}
\end{figure*}

To illustrate the evolution of our model in time for individual discs, in Figure \ref{fig:tracks} we show the evolution of several stars and their corresponding circumstellar discs. These particular tracks are taken from one of the realisations of model R1.0 (see Table \ref{table:models}). We show seven stars with discs as they move through the cluster. Black crosses mark the position of each star at the beginning of the simulation, and the label next to each shows the mass of the star. The sizes of the coloured circles in the stellar tracks are proportional to the disc radii, and their colour indicates the total disc mass. Red crosses, where present, show the moment when the disc is dispersed. The black thin lines that follow a red cross indicate the continuation of the orbit of the star, which keeps moving through the cluster after its disc has been evaporated. The trajectories of some massive, radiating stars are shown in thin blue lines. The solid and dashed circles in the background show the core radius and half mass radius of the cluster, respectively, at $t = \SI{0.0}{Myr}$. A disc around a 0.37 \MSun star survives all through the simulation, however, the variations of its radius in time due to photoevaporation and viscous expansion can be seen. A 0.14 \MSun star initially near the centre keeps its disc until around halfway through the simulation. A very low mass star, 0.06 \MSunns, loses its disc very quickly even if located in the periphery of the cluster. While our simulations are three-dimensional, in this illustrative figure we show a two-dimensional projection of the location of the stars.

\subsection{Disc fractions and lifetimes}\label{results:discfractions}
\begin{figure}
  \includegraphics[width=\linewidth]{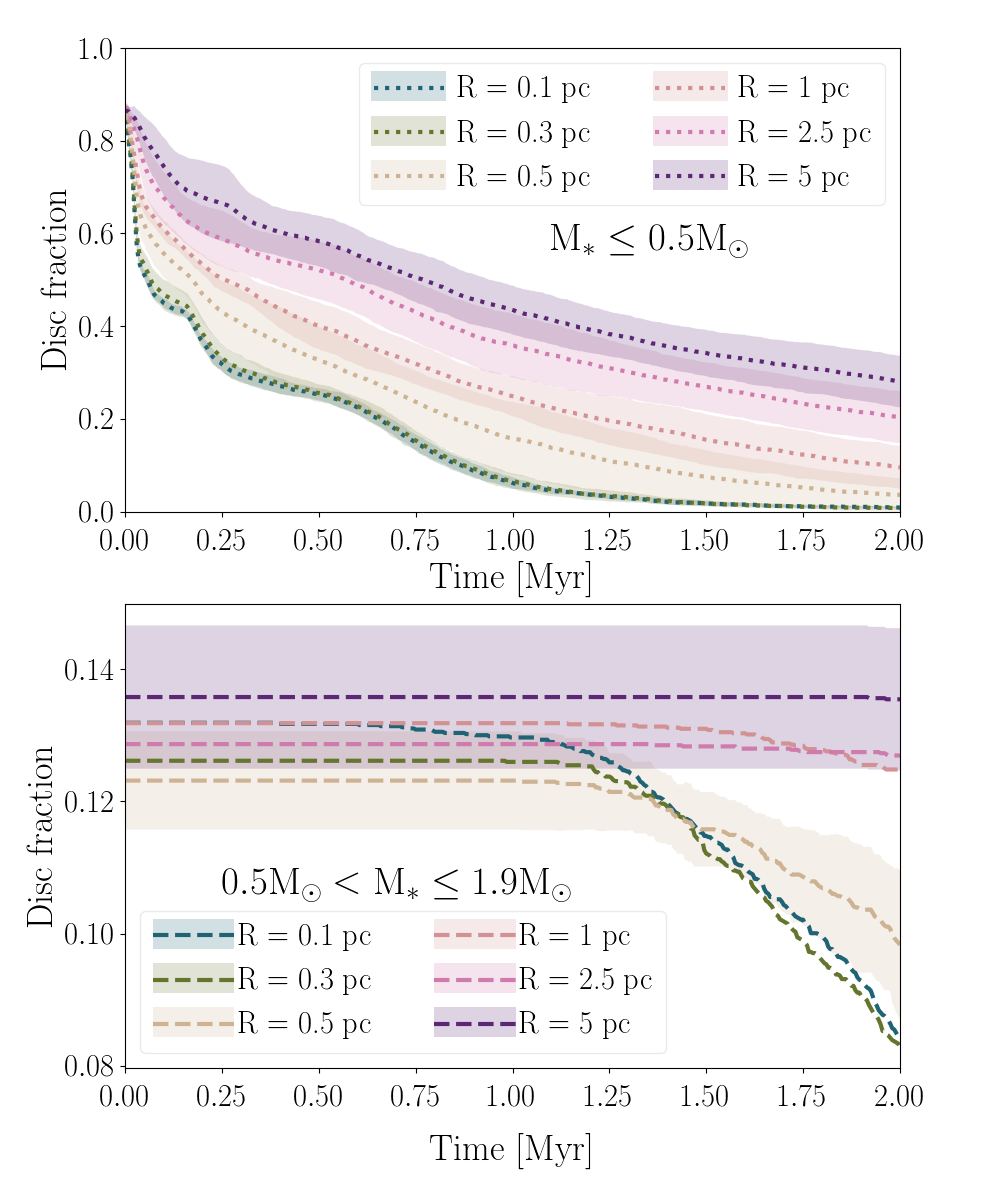}
  \caption{Disc fractions in time separated by stellar mass. The top panel shows disc fractions for low mass stars ($\mathrm{M}_* \leq 0.5 \mathrm{M}_{\odot}$) and the bottom panel for high mass stars ($0.5 \mathrm{M}_{\odot} < \mathrm{M}_* \leq 1.9 \mathrm{M}_{\odot}$). The lines show the mean for 6 runs of each model, and the shaded areas represent the standard deviation. For clarity, in the bottom panel we plot the standard deviation only for the R0.5 and R5.0 models, but the rest of the models have deviations of similar magnitude.}
  \label{fig:separatediscfractions}
\end{figure}

We define the disc fraction at time $t$ as the number of discs at $t$ over the initial number of discs in each cluster. In Figure \ref{fig:separatediscfractions} we show disc fractions separated in terms of the mass of their host stars: low mass stars ($\mathrm{M}_* \leq 0.5 \mathrm{M}_{\odot}$) in the top panel and high mass stars ($0.5 \mathrm{M}_{\odot} < \mathrm{M}_* \leq 1.9 \mathrm{M}_{\odot}$) in the bottom panel. The disc fraction for high mass stars stays constant through time for the R1.0, R2.5, and R5.0 models. These discs lose mass but not enough to be completely evaporated, except for a slight decrease near the end for the R1.0 model. In the R0.1, R0.3, and R0.5 models, however, starting around \SI{1.0}{Myr} even massive discs get destroyed. Final disc fractions decrease with increasing stellar number density. The R0.1 and R0.3 models show a very similar evolution, meaning that the density of the R0.3 model is a higher limit for the effects of external photoevaporation in destroying discs in such simulations.

%\begin{figure*}
%  \includegraphics[width=0.9\linewidth]{figures/disc_fractions.png}
%  \caption{Disc fractions in time. The solid lines show the mean for 6 runs of each model, and the shaded areas represent the standard deviation.}
%  \label{fig:discfractions}
%\end{figure*}

A large drop in the number of stars with discs before \SI{0.2}{Myr} is observed in models R0.1 and R0.3. Similar behaviour was obtained in the simulations performed in Paper I. This drop is related to discs around very low mass stars being dispersed rapidly once the simulation begins and external photoevaporation is `switched on'. This drop can be seen in all the curves, but it becomes less pronounced for lower densities.

In Table \ref{table:lifetimes} we present the mean disc lifetimes and disc half-life for each model, averaged over 6 runs. Disc lifetimes are calculated as a mean of the times when a disc is completely dispersed in the simulations, following the dispersion criteria explained in section \ref{model:photoevaporation}. It is important to mention that this mean is calculated only considering the discs that get dispersed within the $\SI{2.0}{Myr}$ spanned by the simulation, and the discs that survive would likely increase these values. Considering the resulting disc fractions, however, the obtained disc lifetimes are a good approximation. The disc half-life corresponds to the moment when half of the discs in a simulation run have been dispersed. Both the disc lifetimes and the half-life increase with decreasing stellar density.

\begin{table}
\centering
\def\arraystretch{1.5}
\begin{tabular}{ccc} 
\toprule
Model name & Mean disc lifetime [Myr] & Discs half-life [Myr]\\\midrule
R0.1 & $0.38 \pm 0.47$ & $0.20 \pm 0.01$\\
R0.3 & $0.38 \pm 0.47$ & $0.22 \pm 0.03$\\
R0.5 & $0.47 \pm 0.51$ & $0.39 \pm 0.16$ \\
R1.0 & $0.52 \pm 0.55$ & $0.59 \pm 0.11$\\
R2.5 & $0.59 \pm 0.56$ & $0.97 \pm 0.25$\\
R5.0 & $0.65 \pm 0.55$ & $1.42 \pm 0.33$\\
\bottomrule
\end{tabular}\caption{Disc lifetimes and half-life for the different models. First column: model name. Second column: mean disc lifetimes for each model, in Myr. Third column: disc half-life in Myr, calculated as the moment when 50\% of the discs in a simulation have been destroyed. The values are averaged over 6 runs for each model, and the errors represent the variations between runs.}\label{table:lifetimes}
\end{table}

\subsection{Disc masses}\label{results:discmass}
\begin{figure*}
  \includegraphics[width=0.8\linewidth]{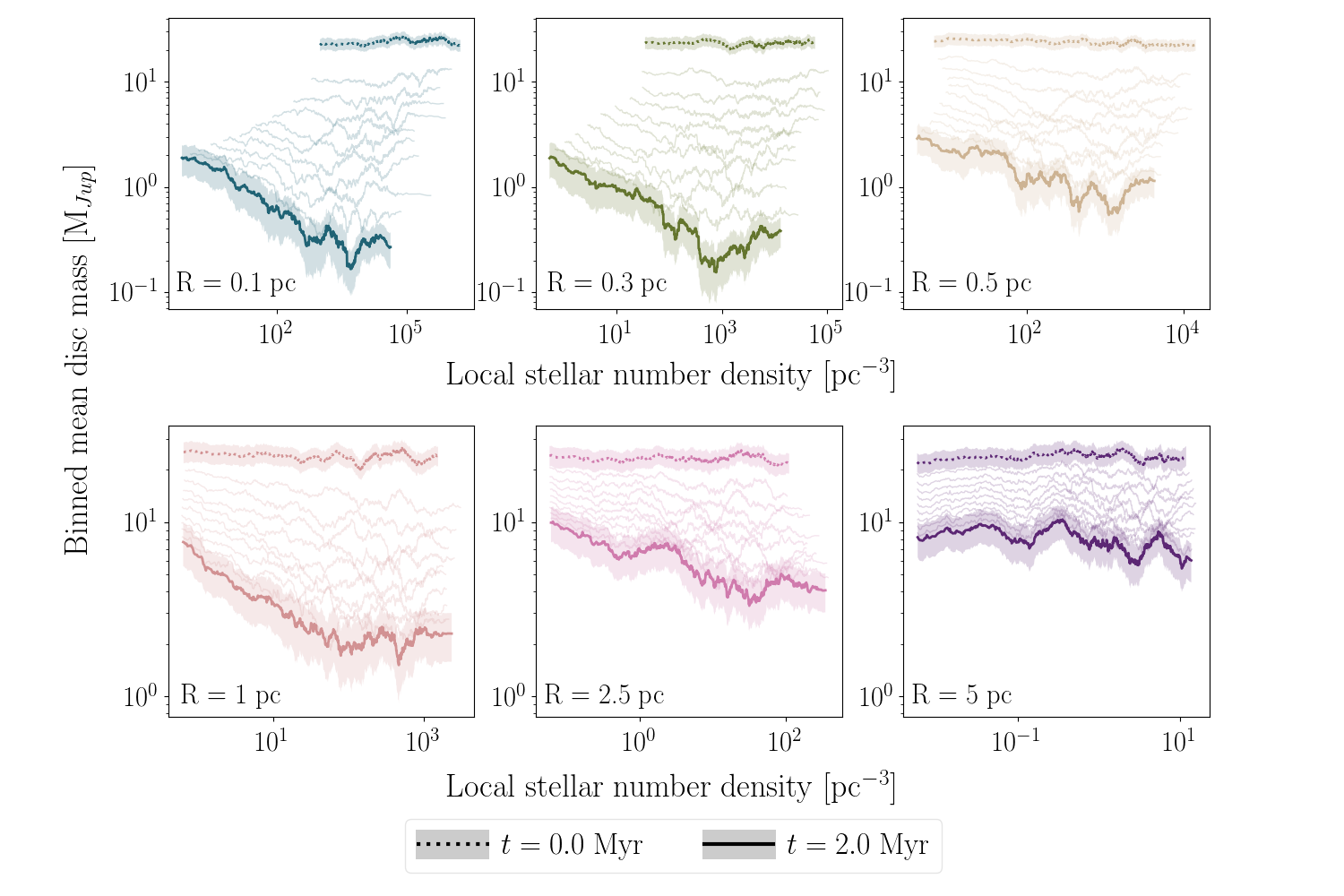}
  \caption{Binned mean disc mass versus local stellar number density. The mean mass is calculated using a moving bin spanning 100 stars. The local density is calculated for each star as explained in section \ref{results:discmass}. The dotted lines thick represent the binned mean disc mass at $t=\SI{0.0}{Myr}$ and the solid thick lines at $t=\SI{2.0}{Myr}$. The shaded areas show the standard error. The thin lines represent the binned mean at $\SI{0.2}{Myr}$ intervals.}
  \label{fig:binnedmean}
\end{figure*}

In Figure \ref{fig:binnedmean} we show the evolution of the mean disc mass in time versus the local stellar number density. The local stellar number density is calculated for each star using the method described by \citet{casertano1985} with the five nearest neighbours. The binned mean disc mass is calculated using a sliding bin spanning 100 stars. 

\begin{figure*}
  \includegraphics[width=0.85\linewidth]{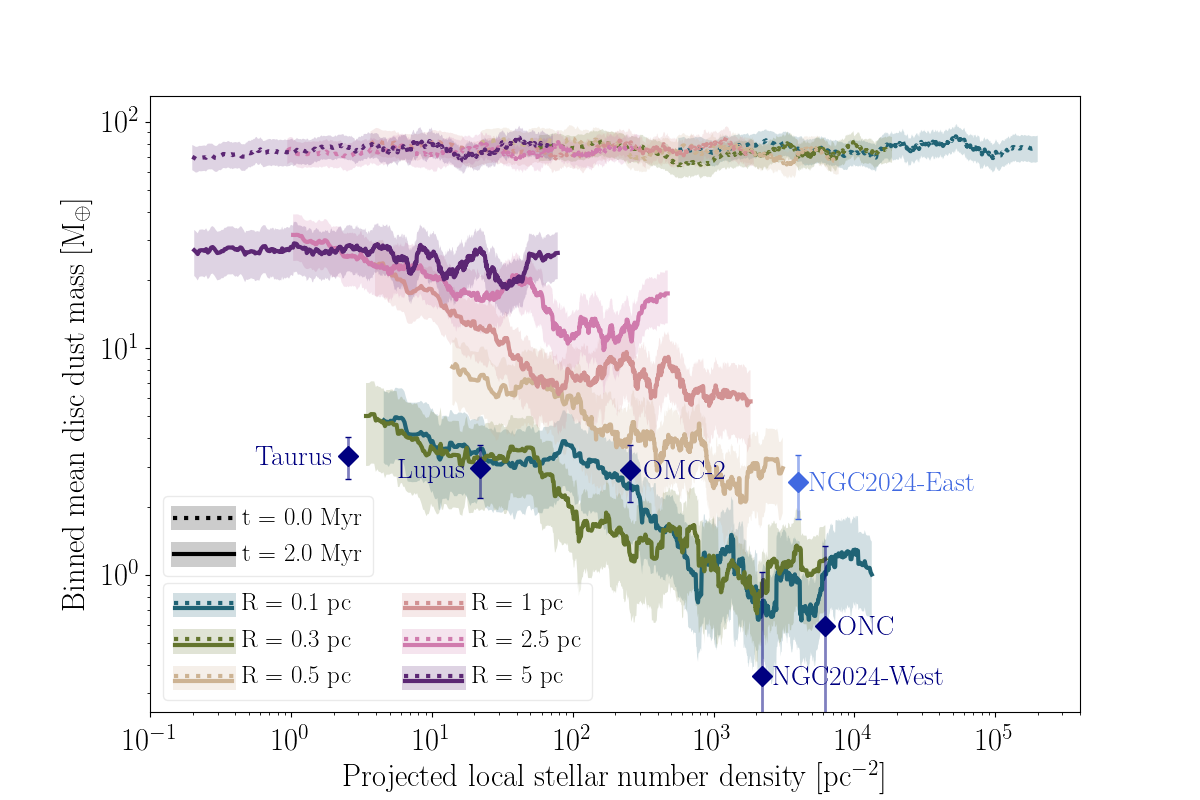}
  \caption{Binned mean disc mass versus local stellar number density, projected in two dimensions. The mean mass is calculated using a moving bin spanning 100 stars. The local density is calculated for each star as explained in section \ref{results:discmass}, but projecting the distances between stars into two dimensions. The dotted lines thick represent the binned mean disc mass at $t=\SI{0.0}{Myr}$ and the solid thick lines at $t=\SI{2.0}{Myr}$. The shaded areas show the standard error. Diamonds show average disc dust masses and local stellar densities for several observed regions. The different color used for NGC 2024 symbolises the different age of the region.}
  \label{fig:allbinnedmeans2D}
\end{figure*}

The thick dotted lines in Figure \ref{fig:binnedmean} show the mean disc mass at $t = \SI{0.0}{Myr}$, and the thick solid lines at $t = \SI{2.0}{Myr}$. The shaded areas around these lines represent the standard error. The thin lines in between show the evolution of the curve in $\SI{0.2}{Myr}$ intervals. The expansion of the clusters in time is reflected by the $t = \SI{2.0}{Myr}$ curves spanning larger density ranges than the $t = \SI{0.0}{Myr}$ curves. This effect is less pronounced in the R1.0, R2.5, and R5.0 models because they expand in a longer time scale. The slope of the final mean disc mass distribution increases with decreasing stellar density. This is related to the core density in each region, which is also decreasing: the curves in the R5.0 model are several orders of magnitude lower, in terms of density, than the R0.1 model. The R0.1 model has a distribution of disc masses such that the most massive discs are found further away from the centre, with differences of about one order of magnitude between the discs located in the centre and in the outskirts of the cluster. In the R5.0 model, the mass difference between discs in different locations is much smaller, and the disc masses are of the same order of magnitude through all the density range.

In Figure \ref{fig:allbinnedmeans2D} we show the mean dust mass of the discs versus the projected local stellar density. We use a 1:100 dust:gas mass ratio to determine the dust mass of our discs. We calculate the density in the same way as in Figure \ref{fig:binnedmean}, but projecting the distances between stars to two dimensions. This allows us to compare disc dust masses in our simulations to observed disc populations. The blue diamonds show points of observed average disc dust mass versus local density of young stellar objects for the Lupus clouds \citep{ansdell2016, ansdell2018}, the Orion Nebula Cluster \citep[ONC,][]{mann2014, eisner2018}, the Orion Molecular Cloud-2 \citep[OMC-2,][]{vanterwisga2019}, Taurus \citep{andrews2013}, and NGC 2024 East and West \citep{getman2014, vanterwisga2020}, as labelled. 

Figure \ref{fig:allbinnedmeans2D} shows a break in disc masses around a local density of 100 stars pc$^{-2}$. The slope of the disc mass distribution changes around that point for all models, except R5.0. In models R2.5 and R1.0 we see the slope of the distributions increasing as we move towards higher densities. For the R0.1, R0.3, and R0.5 models, we see disc mass distributions stay relatively constant for densities lower than 100 stars pc$^{-2}$, and for higher densities we see a negative slope. The difference in masses for discs in regions of density between 100 stars pc$^{-2}$ and $10^4$ stars pc$^{-2}$ is about one order of magnitude. A similar effect can be seen in the observational points, except for NGC 2024 East. This behaviour suggests that 100 stars pc$^{-2}$ is a critical density for determining disc masses.

The average disc dust masses of the observations are calculated by fitting a log-normal distribution on the masses. Although disc masses span a large dynamic range, a log-normal distribution is a good description of these populations \citep{williams2019}. The local density for each point is calculated using the five nearest neighbours method. Lupus data is an average for all the clouds, using the complete list of Class II sources in \citet{ansdell2016} and \citet{ansdell2018}. It is important to note that the Lupus III cloud dominates the signal for that particular region, because it has the largest population of Class II sources. For the OMC-2 the data comes from \citet{vanterwisga2019}, who use the source catalog from \citet{megeath2012} assuming completeness. ONC data comes from \citet{megeath2016}, including completeness corrections. 

In the OMC-2 and ONC regions, observations sample two different density regimes in the same cloud, relatively close together in space. Therefore, the conditions in our models most closely resemble the properties of the discs that were sampled by the observations, and we can interpret them as different density bins along a single model. It is immediately apparent that both the gradient of average disc mass with density as well as the average disc masses themselves resemble the models closely. Given the considerable uncertainties in extracting disc masses from millimeter-continuum observations (see, for instance, the discussion in \citet{eisner2018}) the similarity in the gradients suggests that our models are successful at capturing the general behaviour of external photoevaporation. 

In NGC 2024, \citet{getman2014} find evidence for an age gradient of young stars, which \citet{vanterwisga2020} suggest as an explanation for the large difference in mean disc masses in NGC 2024 East and NGC 2024 West. In NGC 2024, the western part is the older and resembles the ONC in age and conditions, while the eastern disc population has a lower average age. We represent this difference in the figure by making NGC 2024 East in a different shade. Comparing these models to the observations, we see that the NGC 2024 West data lie closely to those of the ONC, while the data for the eastern subpopulation occupy a region of average disc mass and local stellar density space which is more consistent with a younger, compact population of stars.
%It is interesting to note that \citet{getman2014} also suggest such an age gradient in the ONC, albeit less pronounced.

Lupus and Taurus discs, on the other hand, sample a much more heterogeneous part of parameter space in terms of initial densities. Our models do not closely resemble the conditions under which the stars in these samples formed (see, for instance, \citet{roccatagliata2020}). However, the result that when the average stellar number density is low enough (below $\sim$100 stars pc$^{-2}$) the average disc masses are similar at similar ages does seem to apply to these star-forming regions, even though this is a part of parameter space we do not explore.

\section{Discussion}\label{discussion}

%\new{Previous work on theoretical models of external photoevaporation has shown its importance as a process for disc dispersal. In the present work we perform simulations using a self-contained model, in which stellar dynamics, stellar evolution, viscous spreading of the discs, and external photoevaporation are evolved and calculated simultaneously in every time step. This is an improvement over earlier models in which FUV radiation and subsequent mass loss rates were calculated in post-processing \citep[e.g.][]{scally2001, adams2006, fatuzzo2008}.} 

In the simulations performed for this work, external photoevaporation is effective in destroying the majority number of discs within \SI{2.0}{Myr} of evolution. The initial stellar density of each region affects the fraction of surviving discs, as well as their final mass distributions. In all models, except for R5.0, half of the discs are destroyed before \SI{1.0}{Myr} of cluster evolution. A break in the disc mass distributions is seen around 100 stars pc$^{-2}$, in particular for the R0.1 and R0.3 models, with masses dropping about one order of magnitude between 100 stars pc$^{-2}$ and $10^4$ stars pc$^{-2}$. Several surveys have shown that disc mass decreases in the vicinity of bright stars, and in regions with higher stellar number density \citep[e.g.][]{fang2012, mann2009, mann2014, ansdell2017, vanterwisga2020}. In Figure \ref{fig:allbinnedmeans2D} we show our simulation results and compare them with mean disc masses of various observed regions. It is important to note that, due to the nature of estimating disc masses from observations, the mean disc masses from our simulations and from observed regions differ systematically. In observations, mean disc masses are estimated by fitting a log-normal distribution to the measured dust masses \citep[e.g.][]{williams2019}. By calculating the same value for our simulations simply using the mean disc mass, the simulation curves are biased toward high mass disks. Still, the behaviour observed in the simulated disc masses follows the trend of the observations: disc masses decrease as stellar density increases.

The trend in the disc mass distribution for local stellar densities 100 stars pc$^{-2}$ to $10^4$ stars pc$^{-2}$ suggests that, in our models, discs in regions in that density range are less likely to survive long enough or to have enough mass to form planetary systems. The planet formation process should already be underway before \SI{1.0}{Myr} (Figure \ref{fig:separatediscfractions}) for discs to have the minimum reservoir of $10 \mathrm{M}_{\oplus}$ in solids proposed by \citet{ansdell2016} as necessary to form rocky planets or the cores of gas giants. From Figure \ref{fig:allbinnedmeans2D} it can be seen that, in our simulations, all discs in the R5.0, R2.5, and R1.0 models that are in areas of projected local density lower than 100 stars pc$^{-2}$ have masses in excess of $10 \mathrm{\ M}_{\oplus}$. The mean disc dust mass in all other models is below this threshold by \SI{2.0}{Myr}.

Most of the surviving discs in our simulations are around massive stars (0.5 \MSun to 1.9 \MSunns, see Figure \ref{fig:separatediscfractions}). A big factor in this is simply the construction of our models, where initial disc mass is proportional to stellar mass. Figure \ref{fig:separatediscfractions}, however, shows that drops of around 50\% in fractions of discs around high mass stars are still present in high density regions. 

%Following equations \ref{eq:discradius} and \ref{eq:discmass}, these discs around massive stars had initial radii between $\sim\SI{21}{au}$ and $\sim\SI{41}{au}$  and initial masses in the $\sim\SI{52}{\mathrm{M}_{Jup}}$ to $\sim\SI{199}{\mathrm{M}_{Jup}}$ range. Using a 1:100 dust:gas mass ratio we can interpret the initial dust masses of those discs as being between $\sim\SI{166}{\mathrm{M}_{\oplus}}$ and $\sim\SI{6.37E4}{\mathrm{M}_{\oplus}}$. As can be seen in Figure \ref{fig:cumulativemass}, in the R5.0 model this results on disc dust mass distributions between $\sim\SI{5}{\mathrm{M}_\oplus}$ and $\sim\SI{120}{\mathrm{M}_\oplus}$. For the R0.1 model the final dust masses are between $\sim\SI{5}{\mathrm{M}_\oplus}$ and $\sim\SI{100}{\mathrm{M}_\oplus}$, but with around 20\% fewer discs.

At the end of the simulations, the most massive discs are located in areas where the local stellar density is below 100 stars pc$^{-3}$. This implies that large, massive discs observed today either formed in low density regions or migrated to the outskirts of their birth locations fairly quickly. Discs born in the periphery of such regions have a much larger chance of surviving, and we could argue that the disc distributions seen in these low density regions are similar to primordial disc distributions as they are pretty much unperturbed by external photoevaporation.

\section{Conclusions}
We perform simulations of star clusters with circumstellar discs. We implement the stellar dynamics, stellar evolution, viscous evolution of the discs, and external photoevaporation process to evolve simultaneously. We model our clusters as Plummer spheres with $10^3$ stars and initial virial radii of 0.1, 0.3, 0.5, 1.0, 2.5, and $\SI{5.0}{pc}$ to span a range of different number densities. Stars with masses $\mathrm{M}_* \leq 1.9 \mathrm{M}_\odot$ are initially surrounded by a circumstellar disc, and stars with masses $\mathrm{M}_* > 1.9 \mathrm{M}_\odot$ do not have discs and are considered as only emitting UV radiation. Each cluster is evolved for \SI{2.0}{Myr}. We can summarise our findings as follows:

\begin{enumerate}[leftmargin=*]
    \item[1.] External photoevaporation is efficient in destroying circumstellar discs quickly in all simulation models.
    \item[2.] Mean disc lifetimes range from $\SI{0.38 \pm 0.47}{Myr}$ in the denser models ($\mathrm{R}_{\mathrm{cluster}} = [0.1, 0.3, 0.5]$ pc), to $\SI{0.65 \pm 0.55}{Myr}$ for the sparser models ($\mathrm{R}_{\mathrm{cluster}} = [1.0, 2.5, 5.0]$ pc). 
    \item[3.] Disc half-life, the time that it takes for half of the discs to be destroyed in a simulation run, ranges from $\SI{0.20 \pm 0.01}{Myr}$ in the denser models to $\SI{1.42 \pm 0.33}{Myr}$ in the sparser models.
    \item[4.] Disc lifetimes, disc half-lives, disc fractions, and disc masses decrease as the stellar density of the models increase.
    \item[5.] For the final disc masses in the denser regions ($\mathrm{R}_{\mathrm{cluster}} = [0.1, 0.3, 0.5]$ pc) a projected local number density of 100 stars $\mathrm{pc}^{-2}$ introduces a break in the distributions. There are only small variations in the masses of discs around stars in areas of lower densities. As the density increases beyond 100 stars $\mathrm{pc}^{-2}$, the denser regions present a drop of almost an order of magnitude in disc masses.
    %\item[6.] Regions that are initially dense lead to a lower number of discs than in sparser regions, and the surviving discs are less massive. Sparse regions span a larger range of masses of the surviving discs.
    %\item[7.] By the end of our simulations the largest and most massive discs are found in the periphery of the clusters, in particular for the models with higher stellar densities.
    \item[6.] The trends obtained in our simulations between disc mass and local stellar density are in agreement with dust mass measurements of discs in different observed regions: we compare our simulation results to masses of dusty young stellar objects in the Lupus clouds, the Orion Nebula Cluster, the Orion Molecular Cloud-2, Taurus, and NGC 2024.
\end{enumerate}
 
\section*{Acknowledgements}
We would like to thank the anonymous referee for their constructive comments, which greatly helped improve this paper. F.C.-R. would like to thank the Star formation and protoplanetary disc group at Leiden Observatory for helpful discussions and insights. This work was carried out on the Dutch national e-infrastructure with the support of SURF Cooperative. This work was performed using resources provided by the Academic Leiden Interdisciplinary Cluster Environment (ALICE). This paper makes use of the packages \texttt{numpy} \citep{vanderwalt2011a}, \texttt{scipy} \citep{virtanen2019}, \texttt{matplotlib} \citep{hunter2007a}, and \texttt{makecite} \citep{price-whelan2018}.

\section*{Data availability}
The data underlying this article are available at \url{https://doi.org/10.5281/zenodo.3897171}.

%%%%%%%%%%%%%%%%%%%%%%%%%%%%%%%%%%%%%%%%%%%%%%%%%%



\bibliographystyle{mnras}
\bibliography{references} % if your bibtex file is called example.bib

\begin{thebibliography}{}
\makeatletter
\relax
\def\mn@urlcharsother{\let\do\@makeother \do\$\do\&\do\#\do\^\do\_\do\%\do\~}
\def\mn@doi{\begingroup\mn@urlcharsother \@ifnextchar [ {\mn@doi@}
  {\mn@doi@[]}}
\def\mn@doi@[#1]#2{\def\@tempa{#1}\ifx\@tempa\@empty \href
  {http://dx.doi.org/#2} {doi:#2}\else \href {http://dx.doi.org/#2} {#1}\fi
  \endgroup}
\def\mn@eprint#1#2{\mn@eprint@#1:#2::\@nil}
\def\mn@eprint@arXiv#1{\href {http://arxiv.org/abs/#1} {{\tt arXiv:#1}}}
\def\mn@eprint@dblp#1{\href {http://dblp.uni-trier.de/rec/bibtex/#1.xml}
  {dblp:#1}}
\def\mn@eprint@#1:#2:#3:#4\@nil{\def\@tempa {#1}\def\@tempb {#2}\def\@tempc
  {#3}\ifx \@tempc \@empty \let \@tempc \@tempb \let \@tempb \@tempa \fi \ifx
  \@tempb \@empty \def\@tempb {arXiv}\fi \@ifundefined
  {mn@eprint@\@tempb}{\@tempb:\@tempc}{\expandafter \expandafter \csname
  mn@eprint@\@tempb\endcsname \expandafter{\@tempc}}}

\bibitem[\protect\citeauthoryear{Adams, Hollenbach, Laughlin  \& Gorti}{Adams
  et~al.}{2004}]{adams2004}
Adams F.~C.,  Hollenbach D.,  Laughlin G.,   Gorti U.,  2004, \mn@doi [\apj]
  {10.1086/421989}, 611, 360

\bibitem[\protect\citeauthoryear{Adams, Proszkow, Fatuzzo  \& Myers}{Adams
  et~al.}{2006}]{adams2006}
Adams F.~C.,  Proszkow E.~M.,  Fatuzzo M.,   Myers P.~C.,  2006, \mn@doi [\apj]
  {10.1086/500393}, 641, 504

\bibitem[\protect\citeauthoryear{Ali \& Harries}{Ali \&
  Harries}{2019}]{ali2019}
Ali A.~A.,  Harries T.~J.,  2019, \mn@doi [\mnras] {10.1093/mnras/stz1673},
  487, 4890

\bibitem[\protect\citeauthoryear{Anderson, Adams  \& Calvet}{Anderson
  et~al.}{2013}]{anderson2013}
Anderson K.~R.,  Adams F.~C.,   Calvet N.,  2013, \mn@doi [\apj]
  {10.1088/0004-637X/774/1/9}, 774, 9

\bibitem[\protect\citeauthoryear{Andrews, Rosenfeld, Kraus  \& Wilner}{Andrews
  et~al.}{2013}]{andrews2013}
Andrews S.~M.,  Rosenfeld K.~A.,  Kraus A.~L.,   Wilner D.~J.,  2013, \mn@doi
  [\apj] {10.1088/0004-637X/771/2/129}, 771, 129

\bibitem[\protect\citeauthoryear{Ansdell et~al.,}{Ansdell
  et~al.}{2016}]{ansdell2016}
Ansdell M.,  et~al., 2016, \mn@doi [\apj] {10.3847/0004-637X/828/1/46}, 828, 46

\bibitem[\protect\citeauthoryear{Ansdell, Williams, Manara, Miotello, Facchini,
  van~der Marel, Testi  \& van Dishoeck}{Ansdell et~al.}{2017}]{ansdell2017}
Ansdell M.,  Williams J.~P.,  Manara C.~F.,  Miotello A.,  Facchini S.,
  van~der Marel N.,  Testi L.,   van Dishoeck E.~F.,  2017, \mn@doi [\aj]
  {10.3847/1538-3881/aa69c0}, 153, 240

\bibitem[\protect\citeauthoryear{Ansdell et~al.,}{Ansdell
  et~al.}{2018}]{ansdell2018}
Ansdell M.,  et~al., 2018, \mn@doi [\apj] {10.3847/1538-4357/aab890}, 859, 21

\bibitem[\protect\citeauthoryear{Armitage}{Armitage}{2000}]{armitage2000}
Armitage P.~J.,  2000, \aap, 362, 968

\bibitem[\protect\citeauthoryear{Bate}{Bate}{2010}]{bate2010}
Bate M.~R.,  2010, \mn@doi [\mnras] {10.1111/j.1745-3933.2010.00839.x}, 404,
  L79

\bibitem[\protect\citeauthoryear{Bhandare \& Pfalzner}{Bhandare \&
  Pfalzner}{2019}]{bhandare2019}
Bhandare A.,  Pfalzner S.,  2019, \mn@doi [Computational Astrophysics and
  Cosmology] {10.1186/s40668-019-0030-3}, 6, 3

\bibitem[\protect\citeauthoryear{Casertano \& Hut}{Casertano \&
  Hut}{1985}]{casertano1985}
Casertano S.,  Hut P.,  1985, \mn@doi [\apj] {10.1086/163589}, 298, 80

\bibitem[\protect\citeauthoryear{Chevance et~al.,}{Chevance
  et~al.}{2020}]{chevance2020}
Chevance M.,  et~al., 2020, arXiv:2004.06113 [astro-ph]

\bibitem[\protect\citeauthoryear{Clarke}{Clarke}{2007}]{clarke2007}
Clarke C.~J.,  2007, \mn@doi [\mnras] {10.1111/j.1365-2966.2007.11547.x}, 376,
  1350

\bibitem[\protect\citeauthoryear{Close \& Pittard}{Close \&
  Pittard}{2017}]{close2017}
Close J.~L.,  Pittard J.~M.,  2017, \mn@doi [\mnras] {10.1093/mnras/stx897},
  469, 1117

\bibitem[\protect\citeauthoryear{Concha-Ram\'irez, Wilhelm, Portegies~Zwart  \&
  Haworth}{Concha-Ram\'irez et~al.}{2019}]{concha-ramirez2019a}
Concha-Ram\'irez F.,  Wilhelm M. J.~C.,  Portegies~Zwart S.,   Haworth T.~J.,
  2019, \mn@doi [\mnras] {10.1093/mnras/stz2973}, 490, 5678

\bibitem[\protect\citeauthoryear{Eisner \& Carpenter}{Eisner \&
  Carpenter}{2006}]{eisner2006}
Eisner J.~A.,  Carpenter J.~M.,  2006, \mn@doi [\apj] {10.1086/500637}, 641,
  1162

\bibitem[\protect\citeauthoryear{Eisner, Plambeck, Carpenter, Corder, Qi  \&
  Wilner}{Eisner et~al.}{2008}]{eisner2008}
Eisner J.~A.,  Plambeck R.~L.,  Carpenter J.~M.,  Corder S.~A.,  Qi C.,
  Wilner D.,  2008, \mn@doi [The Astrophysical Journal] {10.1086/588524}, 683,
  304

\bibitem[\protect\citeauthoryear{Eisner et~al.,}{Eisner
  et~al.}{2018}]{eisner2018}
Eisner J.~A.,  et~al., 2018, \mn@doi [\apj] {10.3847/1538-4357/aac3e2}, 860, 77

\bibitem[\protect\citeauthoryear{Elmegreen \& Falgarone}{Elmegreen \&
  Falgarone}{1996}]{elmegreen1996}
Elmegreen B.~G.,  Falgarone E.,  1996, \mn@doi [The Astrophysical Journal]
  {10.1086/178009}, 471, 816

\bibitem[\protect\citeauthoryear{Elmegreen, Efremov, Pudritz  \&
  Zinnecker}{Elmegreen et~al.}{2000}]{elmegreen2000}
Elmegreen B.~G.,  Efremov Y.,  Pudritz R.~E.,   Zinnecker H.,  2000, Protostars
  and Planets IV, p.~179

\bibitem[\protect\citeauthoryear{Facchini, Clarke  \& Bisbas}{Facchini
  et~al.}{2016}]{facchini2016}
Facchini S.,  Clarke C.~J.,   Bisbas T.~G.,  2016, \mn@doi [\mnras]
  {10.1093/mnras/stw240}, 457, 3593

\bibitem[\protect\citeauthoryear{Fang et~al.,}{Fang et~al.}{2012}]{fang2012}
Fang M.,  et~al., 2012, \mn@doi [\aap] {10.1051/0004-6361/201015914}, 539, A119

\bibitem[\protect\citeauthoryear{Fatuzzo \& Adams}{Fatuzzo \&
  Adams}{2008}]{fatuzzo2008}
Fatuzzo M.,  Adams F.~C.,  2008, \mn@doi [\apj] {10.1086/527469}, 675, 1361

\bibitem[\protect\citeauthoryear{Font, McCarthy, Johnstone  \& Ballantyne}{Font
  et~al.}{2004}]{font2004}
Font A.~S.,  McCarthy I.~G.,  Johnstone D.,   Ballantyne D.~R.,  2004, \mn@doi
  [\apj] {10.1086/383518}, 607, 890

\bibitem[\protect\citeauthoryear{Getman, Feigelson  \& Kuhn}{Getman
  et~al.}{2014}]{getman2014}
Getman K.~V.,  Feigelson E.~D.,   Kuhn M.~A.,  2014, \mn@doi [\apj]
  {10.1088/0004-637X/787/2/109}, 787, 109

\bibitem[\protect\citeauthoryear{Gorti \& Hollenbach}{Gorti \&
  Hollenbach}{2009}]{gorti2009a}
Gorti U.,  Hollenbach D.,  2009, \mn@doi [\apj] {10.1088/0004-637X/690/2/1539},
  690, 1539

\bibitem[\protect\citeauthoryear{Gorti, Dullemond  \& Hollenbach}{Gorti
  et~al.}{2009}]{gorti2009}
Gorti U.,  Dullemond C.~P.,   Hollenbach D.,  2009, \mn@doi [\apj]
  {10.1088/0004-637X/705/2/1237}, 705, 1237–1251

\bibitem[\protect\citeauthoryear{Guarcello et~al.,}{Guarcello
  et~al.}{2016}]{guarcello2016}
Guarcello M.~G.,  et~al., 2016, arXiv:1605.01773 [astro-ph]

\bibitem[\protect\citeauthoryear{Hacar, Tafalla, Kauffmann  \& Kovács}{Hacar
  et~al.}{2013}]{hacar2013}
Hacar A.,  Tafalla M.,  Kauffmann J.,   Kovács A.,  2013, \mn@doi [Astronomy
  and Astrophysics] {10.1051/0004-6361/201220090}, 554, A55

\bibitem[\protect\citeauthoryear{Haworth \& Clarke}{Haworth \&
  Clarke}{2019}]{haworth2019}
Haworth T.~J.,  Clarke C.~J.,  2019, \mn@doi [\mnras] {10.1093/mnras/stz706},
  485, 3895

\bibitem[\protect\citeauthoryear{Haworth \& Owen}{Haworth \&
  Owen}{2020}]{haworth2020a}
Haworth T.~J.,  Owen J.~E.,  2020, arXiv:2001.05004 [astro-ph]

\bibitem[\protect\citeauthoryear{Haworth, Facchini, Clarke  \& Mohanty}{Haworth
  et~al.}{2018a}]{haworth2018a}
Haworth T.~J.,  Facchini S.,  Clarke C.~J.,   Mohanty S.,  2018a, \mn@doi
  [\mnras] {10.1093/mnras/sty168}, 475, 5460

\bibitem[\protect\citeauthoryear{Haworth, Clarke, Rahman, Winter  \&
  Facchini}{Haworth et~al.}{2018b}]{haworth2018}
Haworth T.~J.,  Clarke C.~J.,  Rahman W.,  Winter A.~J.,   Facchini S.,  2018b,
  \mn@doi [\mnras] {10.1093/mnras/sty2323}, 481, 452

\bibitem[\protect\citeauthoryear{Hollenbach, Yorke  \& Johnstone}{Hollenbach
  et~al.}{2000}]{hollenbach2000}
Hollenbach D.~J.,  Yorke H.~W.,   Johnstone D.,  2000, Protostars and Planets
  IV, p.~401

\bibitem[\protect\citeauthoryear{Hunter}{Hunter}{2007}]{hunter2007a}
Hunter J.~D.,  2007, \mn@doi [Computing in Science & Engineering]
  {10.1109/MCSE.2007.55}, 9, 90

\bibitem[\protect\citeauthoryear{Johnstone, Hollenbach  \& Bally}{Johnstone
  et~al.}{1998}]{johnstone1998}
Johnstone D.,  Hollenbach D.,   Bally J.,  1998, \mn@doi [\apj]
  {10.1086/305658}, 499, 758

\bibitem[\protect\citeauthoryear{Kim, Clarke, Fang  \& Facchini}{Kim
  et~al.}{2016}]{kim2016}
Kim J.~S.,  Clarke C.~J.,  Fang M.,   Facchini S.,  2016, \mn@doi [\apj]
  {10.3847/2041-8205/826/1/L15}, 826, L15

\bibitem[\protect\citeauthoryear{Krause et~al.,}{Krause
  et~al.}{2020}]{krause2020}
Krause M. G.~H.,  et~al., 2020, arXiv:2005.00801 [astro-ph]

\bibitem[\protect\citeauthoryear{Kroupa}{Kroupa}{2001}]{kroupa2001}
Kroupa P.,  2001, \mn@doi [\mnras] {10.1046/j.1365-8711.2001.04022.x}, 322, 231

\bibitem[\protect\citeauthoryear{Krumholz \& Forbes}{Krumholz \&
  Forbes}{2015}]{krumholz2015}
Krumholz M.~R.,  Forbes J.~C.,  2015, \mn@doi [Astronomy and Computing]
  {10.1016/j.ascom.2015.02.005}, 11, 1

\bibitem[\protect\citeauthoryear{Lynden-Bell \& Pringle}{Lynden-Bell \&
  Pringle}{1974}]{lynden-bell1974}
Lynden-Bell D.,  Pringle J.~E.,  1974, \mn@doi [\mnras]
  {10.1093/mnras/168.3.603}, 168, 603

\bibitem[\protect\citeauthoryear{Manara, Morbidelli  \& Guillot}{Manara
  et~al.}{2018}]{manara2018}
Manara C.~F.,  Morbidelli A.,   Guillot T.,  2018, \mn@doi [\aap]
  {10.1051/0004-6361/201834076}, 618, L3

\bibitem[\protect\citeauthoryear{Manara et~al.,}{Manara
  et~al.}{2020}]{manara2020}
Manara C.~F.,  et~al., 2020, arXiv:2004.14232 [astro-ph]

\bibitem[\protect\citeauthoryear{Mann \& Williams}{Mann \&
  Williams}{2009}]{mann2009}
Mann R.~K.,  Williams J.~P.,  2009, \mn@doi [\apjl]
  {10.1088/0004-637X/694/1/L36}, 694, L36

\bibitem[\protect\citeauthoryear{Mann et~al.,}{Mann et~al.}{2014}]{mann2014}
Mann R.~K.,  et~al., 2014, \mn@doi [\apj] {10.1088/0004-637X/784/1/82}, 784, 82

\bibitem[\protect\citeauthoryear{Megeath et~al.,}{Megeath
  et~al.}{2012}]{megeath2012}
Megeath S.~T.,  et~al., 2012, \mn@doi [\apj] {10.1088/0004-6256/144/6/192},
  144, 192

\bibitem[\protect\citeauthoryear{Megeath et~al.,}{Megeath
  et~al.}{2016}]{megeath2016}
Megeath S.~T.,  et~al., 2016, \mn@doi [\apj] {10.3847/0004-6256/151/1/5}, 151,
  5

\bibitem[\protect\citeauthoryear{Nicholson, Parker, Church, Davies, Fearon  \&
  Walton}{Nicholson et~al.}{2019}]{nicholson2019}
Nicholson R.~B.,  Parker R.~J.,  Church R.~P.,  Davies M.~B.,  Fearon N.~M.,
  Walton S. R.~J.,  2019, \mn@doi [\mnras] {10.1093/mnras/stz606}

\bibitem[\protect\citeauthoryear{O'dell}{O'dell}{1998}]{odell1998}
O'dell C.~R.,  1998, \mn@doi [\aj] {10.1086/300178}, 115, 263

\bibitem[\protect\citeauthoryear{O'dell \& Wen}{O'dell \&
  Wen}{1994}]{odell1994}
O'dell C.~R.,  Wen Z.,  1994, \mn@doi [\mnras] {10.1086/174892}, 436, 194

\bibitem[\protect\citeauthoryear{Owen, Ercolano, Clarke  \& Alexander}{Owen
  et~al.}{2010}]{owen2010}
Owen J.~E.,  Ercolano B.,  Clarke C.~J.,   Alexander R.~D.,  2010, \mn@doi
  [\mnras] {10.1111/j.1365-2966.2009.15771.x}, 401, 1415

\bibitem[\protect\citeauthoryear{Pascucci et~al.,}{Pascucci
  et~al.}{2016}]{pascucci2016}
Pascucci I.,  et~al., 2016, \mn@doi [\apj] {10.3847/0004-637X/831/2/125}, 831,
  125

\bibitem[\protect\citeauthoryear{Pelupessy \& Portegies~Zwart}{Pelupessy \&
  Portegies~Zwart}{2012}]{pelupessy2012}
Pelupessy F.~I.,  Portegies~Zwart S.,  2012, \mn@doi [\mnras]
  {10.1111/j.1365-2966.2011.20137.x}, 420, 1503

\bibitem[\protect\citeauthoryear{Pfalzner, Olczak  \& Eckart}{Pfalzner
  et~al.}{2006}]{pfalzner2006}
Pfalzner S.,  Olczak C.,   Eckart A.,  2006, \mn@doi [\aap]
  {10.1051/0004-6361:20064905}, 454, 811

\bibitem[\protect\citeauthoryear{Plummer}{Plummer}{1911}]{plummer1911}
Plummer H.~C.,  1911, \mn@doi [\mnras] {10.1093/mnras/71.5.460}, 71, 460

\bibitem[\protect\citeauthoryear{Portegies~Zwart}{Portegies~Zwart}{2016}]{portegieszwart2016}
Portegies~Zwart S.~F.,  2016, \mn@doi [\mnras] {10.1093/mnras/stv2831}, 457,
  313

\bibitem[\protect\citeauthoryear{Portegies~Zwart \& Verbunt}{Portegies~Zwart \&
  Verbunt}{1996}]{portegieszwart1996}
Portegies~Zwart S.~F.,  Verbunt F.,  1996, \aap, 309, 179

\bibitem[\protect\citeauthoryear{Portegies~Zwart et~al.,}{Portegies~Zwart
  et~al.}{2009}]{portegieszwart2009}
Portegies~Zwart S.,  et~al., 2009, \mn@doi [\na]
  {10.1016/j.newast.2008.10.006}, 14, 369

\bibitem[\protect\citeauthoryear{Portegies~Zwart, McMillan  \&
  Gieles}{Portegies~Zwart et~al.}{2010}]{portegieszwart2010}
Portegies~Zwart S.~F.,  McMillan S. L.~W.,   Gieles M.,  2010, \mn@doi [\araa]
  {10.1146/annurev-astro-081309-130834}, 48, 431

\bibitem[\protect\citeauthoryear{Portegies~Zwart, McMillan, van Elteren,
  Pelupessy  \& de Vries}{Portegies~Zwart et~al.}{2013}]{portegieszwart2013}
Portegies~Zwart S.,  McMillan S. L.~W.,  van Elteren E.,  Pelupessy I.,   de
  Vries N.,  2013, \mn@doi [Computer Physics Communications]
  {10.1016/j.cpc.2012.09.024}, 183, 456

\bibitem[\protect\citeauthoryear{Portegies~Zwart, Pelupessy, van Elteren,
  Wijnen  \& Lugaro}{Portegies~Zwart et~al.}{2018}]{portegieszwart2018}
Portegies~Zwart S.,  Pelupessy I.,  van Elteren A.,  Wijnen T. P.~G.,   Lugaro
  M.,  2018, \mn@doi [\aap] {10.1051/0004-6361/201732060}, 616, A85

\bibitem[\protect\citeauthoryear{Price-Whelan, Mechev  \&
  jumeroag}{Price-Whelan et~al.}{2018}]{price-whelan2018}
Price-Whelan Mechev A.,   jumeroag 2018, adrn/makecite: v0.2, Zenodo,
  \mn@doi{10.5281/zenodo.1343299}

\bibitem[\protect\citeauthoryear{Richling \& Yorke}{Richling \&
  Yorke}{1997}]{richling1997}
Richling S.,  Yorke H.~W.,  1997, \aap, 327, 317

\bibitem[\protect\citeauthoryear{Roccatagliata, Franciosini, Sacco, Randich  \&
  Sicilia-Aguilar}{Roccatagliata et~al.}{2020}]{roccatagliata2020}
Roccatagliata V.,  Franciosini E.,  Sacco G.~G.,  Randich S.,   Sicilia-Aguilar
  A.,  2020, \mn@doi [\aap] {10.1051/0004-6361/201936401}

\bibitem[\protect\citeauthoryear{Rodriguez-Merino, Chavez, Bertone  \&
  Buzzoni}{Rodriguez-Merino et~al.}{2005}]{rodriguez-merino2005}
Rodriguez-Merino L.~H.,  Chavez M.,  Bertone E.,   Buzzoni A.,  2005, \mn@doi
  [\apj] {10.1086/429858}, 626, 411

\bibitem[\protect\citeauthoryear{Scally \& Clarke}{Scally \&
  Clarke}{2001}]{scally2001}
Scally A.,  Clarke C.,  2001, \mn@doi [\mnras]
  {10.1046/j.1365-8711.2001.04274.x}, 325, 449

\bibitem[\protect\citeauthoryear{Scalo}{Scalo}{1990}]{scalo1990}
Scalo J.,  1990, ] {10.1007/978-94-009-0605-1-12}, 162, 151

\bibitem[\protect\citeauthoryear{St{\"o}rzer \& Hollenbach}{St{\"o}rzer \&
  Hollenbach}{1999}]{storzer1999}
St{\"o}rzer H.,  Hollenbach D.,  1999, \mn@doi [\apj] {10.1086/307055}, 515,
  669

\bibitem[\protect\citeauthoryear{Tobin et~al.,}{Tobin et~al.}{2020}]{tobin2020}
Tobin J.~J.,  et~al., 2020, \mn@doi [\apj] {10.3847/1538-4357/ab6f64}, 890, 130

\bibitem[\protect\citeauthoryear{Toonen, Nelemans  \& Zwart}{Toonen
  et~al.}{2012}]{toonen2012}
Toonen S.,  Nelemans G.,   Zwart S.~P.,  2012, \mn@doi [\aap]
  {10.1051/0004-6361/201218966}, 546, A70

\bibitem[\protect\citeauthoryear{Trapman, Rosotti, Bosman, Hogerheijde  \& van
  Dishoeck}{Trapman et~al.}{2020}]{trapman2020}
Trapman L.,  Rosotti G.,  Bosman A.~D.,  Hogerheijde M.~R.,   van Dishoeck
  E.~F.,  2020, arXiv:2005.11330 [astro-ph]

\bibitem[\protect\citeauthoryear{Van Der~Walt, Colbert  \& Varoquaux}{Van
  Der~Walt et~al.}{2011}]{vanderwalt2011a}
Van Der~Walt S.,  Colbert S.~C.,   Varoquaux G.,  2011, \mn@doi [Computing in
  Science & Engineering] {10.1109/MCSE.2011.37}, 13, 22

\bibitem[\protect\citeauthoryear{Vicente \& Alves}{Vicente \&
  Alves}{2005}]{vicente2005}
Vicente S.~M.,  Alves J.,  2005, \mn@doi [\aap] {10.1051/0004-6361:20053540},
  441, 195

\bibitem[\protect\citeauthoryear{Vincke, Breslau  \& Pfalzner}{Vincke
  et~al.}{2015}]{vincke2015}
Vincke K.,  Breslau A.,   Pfalzner S.,  2015, \mn@doi [\aap]
  {10.1051/0004-6361/201425552}, 577, A115

\bibitem[\protect\citeauthoryear{Virtanen et~al.,}{Virtanen
  et~al.}{2019}]{virtanen2019}
Virtanen P.,  et~al., 2019, arXiv:1907.10121 [physics]

\bibitem[\protect\citeauthoryear{Wijnen, Pols, Pelupessy  \&
  Portegies~Zwart}{Wijnen et~al.}{2016}]{wijnen2016}
Wijnen T. P.~G.,  Pols O.~R.,  Pelupessy F.~I.,   Portegies~Zwart S.,  2016,
  \mn@doi [\aap] {10.1051/0004-6361/201527886}, 594, A30

\bibitem[\protect\citeauthoryear{Wijnen, Pols, Pelupessy  \&
  Portegies~Zwart}{Wijnen et~al.}{2017}]{wijnen2017}
Wijnen T. P.~G.,  Pols O.~R.,  Pelupessy F.~I.,   Portegies~Zwart S.,  2017,
  \mn@doi [\aap] {10.1051/0004-6361/201630221}, 602, A52

\bibitem[\protect\citeauthoryear{Williams \& Best}{Williams \&
  Best}{2014}]{williams2014}
Williams J.~P.,  Best W. M.~J.,  2014, \mn@doi [\apj]
  {10.1088/0004-637X/788/1/59}, 788, 59

\bibitem[\protect\citeauthoryear{Williams \& Cieza}{Williams \&
  Cieza}{2011}]{williams2011}
Williams J.~P.,  Cieza L.~A.,  2011, \mn@doi [\araa]
  {10.1146/annurev-astro-081710-102548}, 49, 67

\bibitem[\protect\citeauthoryear{Williams, Cieza, Hales, Ansdell,
  Ruiz-Rodriguez, Casassus, Perez  \& Zurlo}{Williams
  et~al.}{2019}]{williams2019}
Williams J.~P.,  Cieza L.,  Hales A.,  Ansdell M.,  Ruiz-Rodriguez D.,
  Casassus S.,  Perez S.,   Zurlo A.,  2019, arXiv:1904.06471 [astro-ph]

\bibitem[\protect\citeauthoryear{Winter, Clarke, Rosotti, Ih, Facchini  \&
  Haworth}{Winter et~al.}{2018}]{winter2018a}
Winter A.~J.,  Clarke C.~J.,  Rosotti G.,  Ih J.,  Facchini S.,   Haworth
  T.~J.,  2018, \mn@doi [\mnras] {10.1093/mnras/sty984}, 478, 2700

\bibitem[\protect\citeauthoryear{Winter, Clarke, Rosotti, Hacar  \&
  Alexander}{Winter et~al.}{2019}]{winter2019a}
Winter A.~J.,  Clarke C.~J.,  Rosotti G.~P.,  Hacar A.,   Alexander R.,  2019,
  \mn@doi [\mnras] {10.1093/mnras/stz2545}, 490, 5478

\bibitem[\protect\citeauthoryear{Winter, Ansdell, Haworth  \& Kruijssen}{Winter
  et~al.}{2020a}]{winter2020}
Winter A.~J.,  Ansdell M.,  Haworth T.~J.,   Kruijssen J. M.~D.,  2020a,
  \mn@doi [\mnrasl] {10.1093/mnrasl/slaa110}, p. slaa110

\bibitem[\protect\citeauthoryear{Winter, Kruijssen, Chevance, Keller  \&
  Longmore}{Winter et~al.}{2020b}]{winter2020a}
Winter A.~J.,  Kruijssen J. M.~D.,  Chevance M.,  Keller B.~W.,   Longmore
  S.~N.,  2020b, \mn@doi [\mnras] {10.1093/mnras/stz2747}, 491, 903

\bibitem[\protect\citeauthoryear{van Terwisga, Hacar  \& van Dishoeck}{van
  Terwisga et~al.}{2019}]{vanterwisga2019}
van Terwisga S.~E.,  Hacar A.,   van Dishoeck E.~F.,  2019, \mn@doi [\aap]
  {10.1051/0004-6361/201935378}, 628, A85

\bibitem[\protect\citeauthoryear{van Terwisga et~al.,}{van Terwisga
  et~al.}{2020}]{vanterwisga2020}
van Terwisga S.~E.,  et~al., 2020, arXiv:2004.13551 [astro-ph]

\makeatother
\end{thebibliography}

% Don't change these lines
\bsp	% typesetting comment
\label{lastpage}
\end{document}